# Spherical and Rod-shaped Gold Nanoparticles for Surface Enhanced Raman Spectroscopy

Md. Shaha Alam[*†], Syed Farid Uddin Farhad[*§†], Nazmul Islam Tanvir [*§†], Md. Nur Amin Bitu [§†], Mohammad Moniruzzaman[*†], Mahmuda Hakim[†], and Md. Aftab Ali Shaikh[†]

[*]Central Analytical and Research Facilities (CARF), Dhaka 1205, Bangladesh
[§]Energy Conversion and Storage Research, Industrial Physics Division, BCSIR Labs, Dhaka 1205, Bangladesh
[†]Bangladesh Council of Scientific and Industrial Research (BCSIR), Bangladesh
Email: s.f.u.farhad@bcsir.gov.bd

*Abstract*— Raman Spectroscopy offers an in-situ, rapid, and non-destructive characterization tool for chemical analysis of diverse samples with no or minimal preparation. However, due to the inherent weak signal of conventional Raman spectroscopy, surface plasmon resonance features of noble metal nanoparticles have been utilized to conduct Surface Enhanced Raman Spectroscopy (SERS) in detecting trace label contaminants in foods and foodstuffs. In this effort, we synthesized gold nanoparticles (AuNPs) by reduction of chloroauric acid ($HAuCl_4$) with sodium citrate dehydrate. We prepared different sizes of AuNPs at a fixed temperature (100 ºC) but with varying pHs of 4 and 8. The as-synthesized AuNPs were characterized by UV–Vis spectroscopy, dynamic light scattering (DLS), and Field Emission Scanning Electron Microscopy (FE-SEM). FE-SEM micrographs revealed spherical AuNPs with an average diameter of ~ 55 ± 13 nm and rod-shaped AuNPs with an average length of ~ 170 ± 36 nm for samples synthesized at pH 8 and 4, respectively. The effectiveness of the as-prepared AuNPs for SERS is tested by detecting Rhodamine 6G diluted at a trace level. This study suggests that plasmonic nanoparticles coupled with SERS have great potential for broad applications in detecting other trace amounts of hazardous chemicals in foods and foodstuffs.

*Keywords*— *Gold nanoparticles, Surface Enhanced Raman Scattering, Rhodamine 6G, Liquid phase SERS*

## I. Introduction

Raman spectroscopy is a simple but powerful technique to determine the vibrational structure of molecules with no or minimal sample preparation [1-3]. Due to its non-destructive sample analysis nature, Raman spectroscopy has been widely used in many disciplines, including analytical chemistry [2,3], materials science [4-7], food and agricultural industries ([3] and refs. therein), biological science [8,9], as well as forensic and pharmaceutical science [10,11]. However, due to the inherent weak signal of conventional Raman spectroscopy, surface plasmon resonance features of noble metal nanoparticles have been utilized to increase the strength of Raman signals. The latter procedure is called Surface Enhanced Raman Scattering (SERS) [1-3, 12]. Among various nanomaterials, gold nanoparticles (AuNPs) draw significant attention to scientists in SERS measurements [8-13]. For SERS enhancement, the commercial substrates are expensive, and preparation methods are complicated and time-consuming. To address this issue, we synthesized different sizes of AuNPs using a facile technique and checked their utility by enhancing the Raman intensity of Rhodamine 6G (R6G). R6G is a cationic dye toxic to living organisms, cause skin allergic reactions, irritates the respiratory tract, and affects the central nervous system, liver, reproductive system, eyes, etc.[14]. Food and Drug Administration (FDA) banned its use in food additives; however, due to its low-cost many food industries use it as a food colorant [15-18]. In this study, we prepared AuNPs of different sizes and shapes to achieve the best SERS enhancement of R6G diluted sequentially at a trace level.

## II. Experimental Section

### A. Chemicals

Auric chloride ($HAuCl_4$) (purity~99.99 %) and sodium citrate dehydrate (purity~99.0 %) were purchased from Sigma-Aldrich and used without further purification. Hydrochloric acid (certified ACS Plus) and nitric acid (certified ACS Plus) were purchased from Fisher Scientific. Deionized water (18.2 MΩ-cm) was used for all solution preparations.

### B. Synthesis of gold nanoparticles

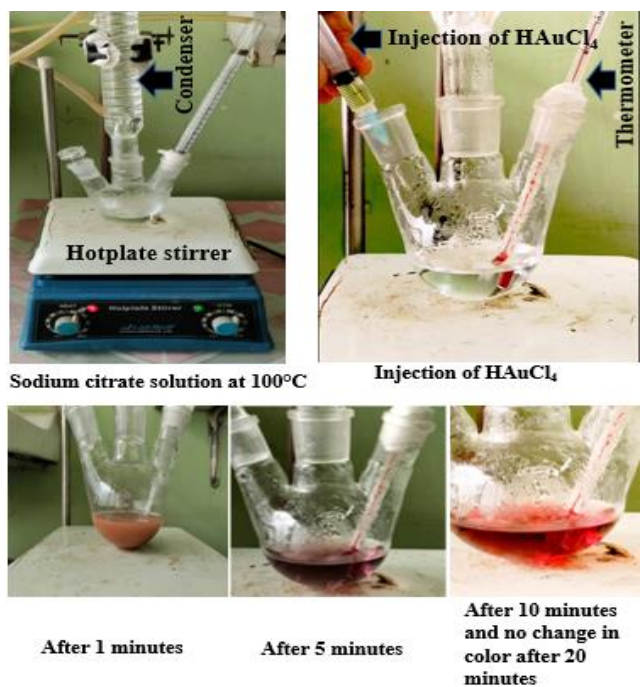

Fig. 1. Synthesis procedure of gold nanoparticles (AuNPs)

In a typical procedure of AuNPs, 30.0 mL of 19.4 mM sodium citrate dihydrate solution (adjusted pH at 8.0 by adding 5 M NaOH) and a magnetic stir bar were added to a three-angled neck flask. A condenser was fixed to the flask's center neck, a thermometer and a glass stopper were put on the other two sides' neck, then placed on a hotplate stirrer. The solution was boiled for 20 min at 100 $^0$C, and then freshly prepared 1.0 mL of 5 mM $HAuCl_4$ solution was added. The reaction was allowed to continue until the wine-red color was observed. The wine-red color indicates the formation of gold nanoparticles (AuNPs). The same procedure was also followed at pH 4 (adjusted by adding 5 M HCl) for 10 mM sodium citrate dihydrate solution and 1 mM $HAuCl_4$ to control the shape and size of the resulting AuNPs (solution color was violet). The graphical presentation of the whole synthesis procedure is shown in Fig. 1.

*C. Characterizations and Instrumentations*

The surface plasmon resonance (SPR) bands of different AuNPs were recorded at room temperature (~25 °C) using Shimadzu 2600 UV-VIS-NIR spectrometer. The maximum surface plasmon resonance (SPR) wavelength ($\lambda_{max}$) of the resulting AuNPs solution was calculated from the observed UV-Vis spectrum. Dynamic Light Scattering (DLS) (Horiba SZ-100V2) has been used for average particle size (APS), size distribution, and zeta potential of AuNPs solution. The surface morphology, size, and shape of the AuNPs were examined by a Field Emission-Scanning Electron Microscope (FE-SEM, JOEL 7610F). The SERS studies of AuNPs coupled with R6G were recorded at room temperature using a Raman spectrometer (Horiba MacroRam) with a 785 nm diode laser as the excitation source (Laser power kept < 5 mW at the sample surface to minimize the sample degradation).

## III. RESULTS AND DISCUSSION

The color change in the reaction solution from pale yellow to wine red (Fig. 1) indicates that the AuNPs formation is completed, and one can easily spot the difference in color change. The sodium citrate functions as a stabilizing agent, a capping agent, and a reducing agent to turn $Au^{+3}$ ion to $Au^0$. Intriguingly, violet and red-wine AuNPs solutions were found at pH 4 and 8 of the precursors.

Fig. 2(a, b) and their respective inset represent the UV–Vis spectra of as-synthesized AuNPs and solution color. Interestingly, AuNPs are seen to have a specific surface plasma resonance (SPR) band between 450 to 600 nm, red-wine AuNPs solution with $\lambda_{max} \approx 534$ nm (Fig.2a) and violet AuNPs solution with $\lambda_{max} \approx 542$ nm (Fig.2b) due to size and shape variation of nanoparticles in the solutions under study. Larger particles underwent increased scattering and had a resonance peak that broadened significantly and shifted towards longer wavelengths (known as red-shift). This was further confirmed by the DLS measurements: APS ≈ 50 nm for $\lambda_{max} \approx 534$ nm AuNPs solution and APS ≈ 193 nm for $\lambda_{max} \approx 542$ nm AuNPs solution shown in Fig. 3. The smaller AuNPs solution was found to have zeta potential (mean) is -34.4 mV compared to -12.9 mV for larger AuNPs solution. These UV-Vis and DLS observations are in good agreement with the studies reported by others [19-21].

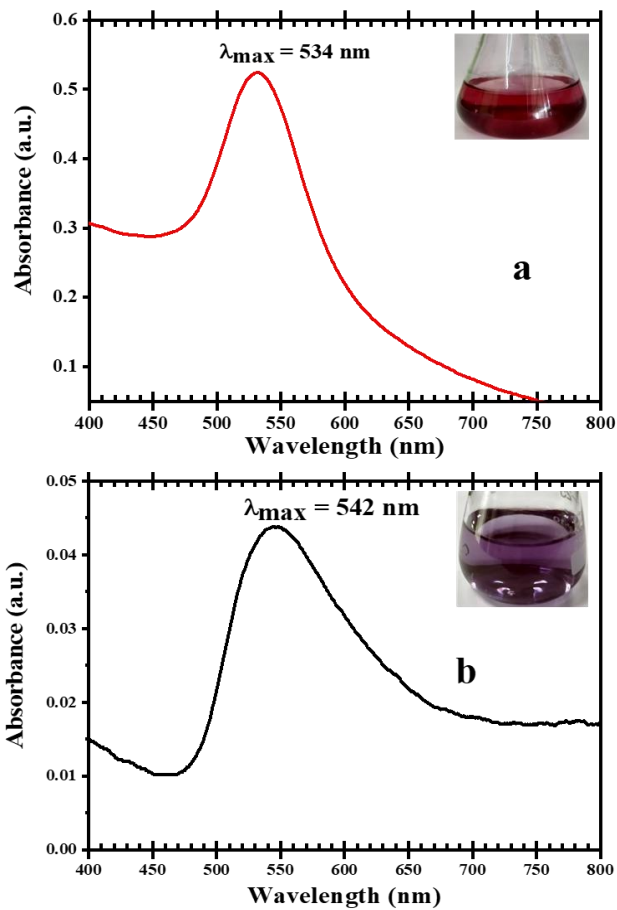

Fig. 2. UV-Vis spectrum of as-synthesized AuNPs

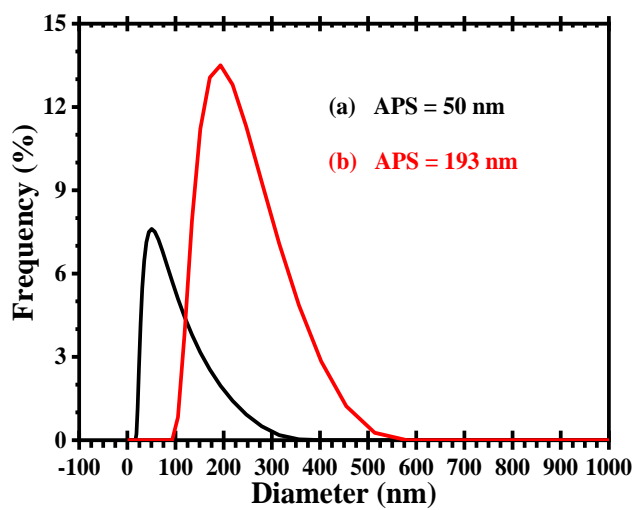

Fig. 3. DLS graph shows the average particle size (APS) and size distribution of AuNPs synthesized under two conditions.

The size and shape of AuNPs may affect the SERS enhancement of samples under study [22, 23]. The morphology of AuNPs affects their attachment to the sample under study (e.g., R6G). To this end, FE-SEM micrographs of AuNPs grown at pH 8 and pH 4 of precursor solution were analyzed by ImageJ software [7], and the results are displayed with the respective image in Fig. 4.

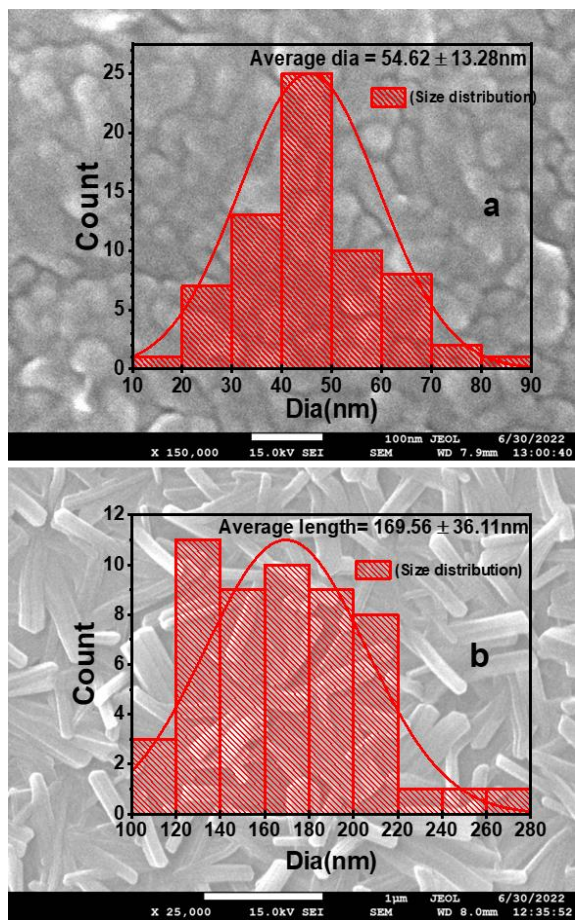

Fig. 4: Size distribution curve of AuNPs curve from FE-SEM

As can be seen from Fig. 4, FE-SEM images revealed almost spherical-shaped AuNPs with an average diameter of ~ 55 ± 13 nm for red-wine solution (Fig. 4a) and rod-shaped AuNPs with an average length of ~ 170 ± 36 nm for violet solutions. Indeed, the FE-SEM analyzed size measurements corroborate the DLS-derived size measurements within the error margin.

*SERS spectra of Rhodamine 6G (R6G)*

To elucidate the effect of spherical and rod-shaped AuNPs on the vibrational structure of R6G (0.2 mM), SERS of R6G adsorbed on colloidal AuNPs were studied in the range of 180 - 2000 $cm^{-1}$. SERS samples were prepared by drop casting of 5-10 μL AuNPs on R6G droplets on the stainless steel (SS) substrates, and then Raman spectra were recorded immediately on the liquid phase sample (without drying). To distinguish the Raman signature of bare SS substrate, AuNPs, and R6G were also recorded under the same conditions and displayed in Fig. 5. As can be seen from this figure, bare SS (Fig.5a), AuNPs on SS (Fig. 5b), and, R6G on SS (Fig. 5c) did not exhibit any sharp Raman peaks in the range of 500 - 1800 $cm^{-1}$, except a broad hump near ~1400 $cm^{-1}$. In the case of AuNPs on SS sample, the vibration peak near ~ 250 $cm^{-1}$ was presumably induced from the plasmonic features of AuNPs, which is absent in bare SS, and R6G on SS but present again in AuNP (50 nm) on R6G (Fig. 5d) and AuNP (193 nm) on R6G (Fig.5e) samples. Intriguingly, 50 nm AuNP (spherical) coupled R6G produced significant SERS enhancement compared to that of 193 nm AuNP (nanorod) associated R6G sample (cf. Fig. 5d and Fig. 5e). The SERS enhanced Raman peaks of R6G are in good agreement with our previous report done in solid phase sample [24].

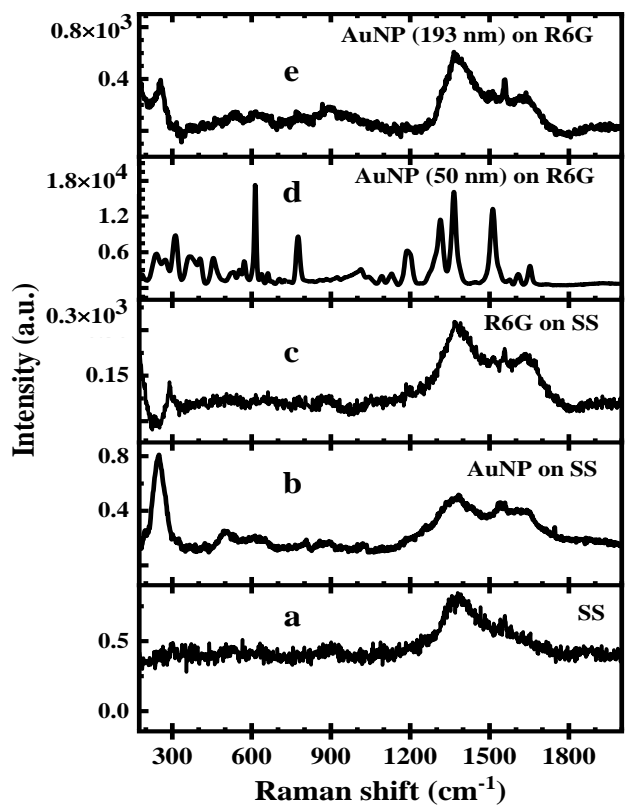

Fig. 5: Raman spectrum of (a) stainless steel substrate (SS), (b) AuNP on SS, (c) R6G on SS, (d, e) AuNP on R6G.

*Effect of AuNP size and shape on SERS enhancement*

Surface plasmon resonance (SPR) produced by incident laser beam enhanced the electromagnetic field in the vicinity of nanoparticles, which play a crucial role in the Raman signals of the nanoparticle's adjacent molecules. Nanoparticle size, shape, and degree of their aggregation affect the SERS enhancement [25]. Spherical-shaped and smaller size (~50 nm) AuNPs exhibited the highest SERS enhancement compared to rod-shaped and larger size (~193 nm) AuNPs corroborating the reported results [26]. The present study was done on diluted 0.2 mM R6G, which suggests that plasmonic nanoparticles coupled with SERS have great potential for broad applications in detecting other trace amounts of hazardous chemicals in foods and foodstuffs. Further investigation of different sizes and shapes of AuNPs is currently in progress to elucidate the critical plasmonic size of AuNPs for detecting trace levels of pesticides added to food and agricultural products.

IV. CONCLUSION

This report describes a facile technique to synthesize size and shape-controlled gold nanoparticles (AuNPs) by controlling the pH and concentration of precursor solutions.

As-synthesized AuNPs were characterized by various characterization tools, including UV-Vis Spectroscopy, DLS, and FE-SEM, to elucidate their size and shape effect on the SERS activity of diluted Rhodamine 6G. This study recapitulates that it provides a simple, rapid, and sensitive approach to analyzing trace amounts of hazardous chemicals to ensure food safety.

ACKNOWLEDGMENT

All the authors gratefully acknowledge the experimental support of the Energy Conversion and Storage Research (ECSR) Section, Industrial Physics Division (IPD), BCSIR Laboratories, Dhaka 1205, Bangladesh Council of Scientific and Industrial Research (BCSIR), under the scope of R&D project#TCS-FY2017-2022. S.F.U.F. and N.I.T. acknowledge the support of Special Allocation Grant#404-ES-FY2021-22(Ref.#39.00.0000.009.14.019.21-745 Dated 15/12/2021), Ministry of Science and Technology, Government of Bangladesh, TWAS Grant#20-143 RG/PHYS/AS_I, and RSC(UK) Grant# R20-3167 for ECSR, IPD.